\documentstyle[amssymb,multicol,epsfig,rotate,aps]{revtex}

\begin{document}
\draft
 
\title{Multiple scaling regimes in simple aging models}

\author{Bernd Rinn,$^1$ Philipp Maass,$^{1,2}$ and Jean-Philippe Bouchaud$^2$}

\address{$^1$Fakult\"at f\"ur Physik, Universit\"at Konstanz, D-78457
  Konstanz, Germany} \address{$^2$Service de Physique de l'Etat
  Condens\'e, CEA Saclay, 91191 Gif sur Yvette Cedex, France}

\date{January 12, 2000}

\maketitle

\begin{abstract} 
  We investigate aging in glassy systems based on a simple model,
  where a point in configuration space performs thermally activated
  jumps between the minima of a random energy landscape. The model
  allows us to show explicitly a subaging behavior and multiple
  scaling regimes for the correlation function. Both the exponents
  characterizing the scaling of the different relaxation times with
  the waiting time and those characterizing the asymptotic decay of
  the scaling functions are obtained analytically by invoking a
  `partial equilibrium' concept.
\end{abstract}

% 02.50.-r Probability theory, stochastic processes, and statistics
% 75.10.Nr Spin-glass and other random models
% 05.20.-y Classical statistical mechanics

\begin{multicols}{2}
\narrowtext

The dynamics of glassy materials can be strongly dependent on the
history of glass formation \cite{Zarzycki:1991,Parisi:1999}. Generally
speaking, one finds that the relaxation dynamics becomes increasingly
slower with the ``age'' of the system, that means with the time $t_w$
expired since the material was brought into the glassy state. Such
aging phenomena have been identified in many systems and various
dynamical probes (for a recent review, see
e.g.~\cite{Bouchaud/etal:1998}).  Prominent examples are shear stress
relaxations in structural glasses \cite{Struik:1978}, thermoremanent
magnetizations in spin glasses \cite{Vincent/etal:1997}, and electric
field relaxations in dipolar glasses \cite{Alberici/etal:1997}. A
convenient way to quantify aging in such experiments is to disturb the
probe at time $t_w$ by a sudden change of external field, and to
measure the response $R(t_w+t,t_w)$ at a later time $t_w+t$. Often, the
characteristic relaxation time grows proportionally to the age 
$t_w$. When $t$ is larger than all microscopic times associated
with fast time-translational invariant relaxations, one then expects
$R(t_w+t,t_w)$ to depend on the ratio $t/t_w$ only, i.e.
$R(t_w\!+\!t,t_w)=F(t/t_w)$.

In principle, however, one cannot rule out other scaling forms, as
e.g.  $R(t_w\!+\!t,t_w)=F(t/t_w^\mu)$ with $\mu>0$ being different
from one.  In particular the case $\mu<1$, which has been called
`subaging' because the effective relaxation time grows more slowly
than the age of the system, seems to be of experimental relevance
\cite{Bouchaud/etal:1998}. It is moreover possible that there exist,
for given waiting time $t_w$, various scaling regimes in time $t$,
which are governed by {\it different} relaxation times $\propto
t_w^{\mu_s}$, $s=1,2,\ldots$. More precisely, depending on how $t$ is
scaled with $t_w$, one can obtain different asymptotic scaling
functions in the limit $t_w\!\to\!\infty$. For example, for
$\mu_1\!>\!\mu_2\!>\!0$, one may find
$R(t_w\!+\!\Lambda_1\,t_w^{\mu_1},t_w)\!\sim\!F_1(\Lambda_1)$ and
$R(t_w\!+\!\Lambda_2\,t_w^{\mu_2},t_w)\!\sim\!F_2(\Lambda_2)$ when
$t_w\to\infty$. In fact, the occurrence of different scaling functions
being associated with various time regimes has recently been
conjectured on the basis of analytical results for the Langevin
dynamics of mean-field spin glass models
\cite{Cugliandolo/Kurchan:1994,Bouchaud/etal:1998}. So far, however,
it was not possible to validate these conjectures, or to exemplify
them in some reasonable phenomenological models.

In this Letter we will discuss a model that allows us to demonstrate
for the first time explicitly the possible occurrence of subaging
behavior and multiple scaling regimes. This model, which has a strong
resemblance to the previously studied ``trap model''
\cite{Bouchaud/Dean:1995,Feigelman/etal:1988}, is motivated by the
simple and widespread view that glassy dynamics may be described by a
thermally activated motion of a point (``particle'') that jumps among
the deep (free) energy minima $E_i$ of a complex configuration space.
According to extreme value statistics one may expect the distribution
$\rho(E)$ of these deep minima to be exponential, and indeed,
mean-field theories of spin glasses \cite{Bouchaud/Mezard:1997} and
recent results from molecular dynamics simulations
\cite{Schoen/Sibani:1999} suggest this to be the case.

To be specific, let us consider a $d$-dimensional cubic lattice and
assign to each lattice site $i$ an energy $E_i$,
$-\infty\!<\!E_i\!\le\!0$, drawn from the distribution
$\rho(E)\!=\!T_{\rm g}^{-1}\exp(E/T_{\rm g})$.  The particle jumps among
nearest neighbor sites only, and the jump rate from site $i$ to a
neighboring site $j$ is
\begin{equation}
w_{i,j}=\nu\exp\left(-\beta[\alpha E_j-(1-\alpha)E_i]\right)\,,
\label{wij-eq}
\end{equation}
where the ``attempt frequency'' $\nu\!\equiv\!1$ sets our time
unit, $\beta^{-1}\!\equiv\!T$ is the temperature (or thermal energy),
and the parameter $\alpha$ specifies how the energies of the initial
and target site are weighted. In order for the $w_{i,j}$ to obey
detailed balance, $\alpha$ can assume any real value, but on physical
grounds it is reasonable to restrict $\alpha$ to the range
$0\!\le\!\alpha\!\le\!1$. Independent of $\alpha$, the system
undergoes a ``dynamical phase transition'' at $T\!=\!T_{\rm g}$: In
the high-temperature phase, where $T\!>\!T_{\rm g}$, the probability
$\varphi(E)\!\propto\!\rho(E)\exp(-\beta E)$ for finding the system in
a state with energy $E$ is normalizable and thermal equilibrium will
be approached with a characteristic equilibration time that diverges
for $T\!\searrow\!T_{\rm g}$. By contrast, in the glassy phase, where
$T\!<\!T_{\rm g}$ and $\varphi(E)$ is not normalizable, the dynamics never
becomes stationary but `ages'.

It is important at this point to stress the differences between the
above model and the earlier studied trap model
\cite{Bouchaud/Dean:1995}. In the latter, the jump rates depend only
on the energy of the initial site corresponding to $\alpha\!=\!0$ in
eq.~(\ref{wij-eq}), and this allows a straightforward mapping onto a
continuous time-random walk with a waiting time distribution decaying
as a power law. Much more important,
%************************************************************************
\begin{figure}[t!]
\hspace*{-0.5cm}\epsfig{file=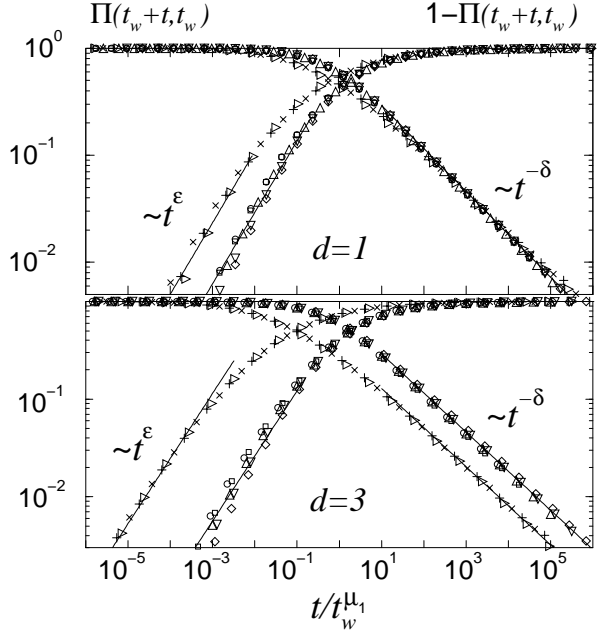,width=8.5cm}
\caption{Double-logarithmic plot of $\Pi(t_w\!+\!t,t_w)$,
  $1-$ $\Pi(t_w\!+\!t,t_w)$ (symbols
  $\circ,\,{\scriptstyle\square},\,{\scriptstyle\triangle},\,\triangledown,\,
  \diamond$) and $\tilde\Pi(t_w\!+\!t,t_w)$,
  $1\!-\!\tilde\Pi(t_w\!+\!t,t_w)$ (symbols
  $\triangleright,\,+,\,\times$) as functions of $t/t_w^{\mu_1}$ in
  $d\!=\!1$ and $d\!=\!3$ for $(\theta,\alpha)\!=\!(1/4,3/8)$.
  Different symbols correspond to different waiting times,
  $t_w\!=\!10^5$ ($\diamond$), $10^6$ ($\triangledown$), $10^7$
  (${\scriptstyle\triangle}$), $10^8$ (${\scriptstyle\square}$),
  $10^9$ ($\circ$) and $t_w\!=\!10^8$ ($\triangleright$), $10^{11}$
  ($+$), $10^{13}$ ($\times$) in $d=1$, as well as $t_w\!=\!10^6$
  ($\diamond$), $10^7$ ($\triangledown$), $10^8$
  (${\scriptstyle\triangle}$), $10^9$ (${\scriptstyle\square}$),
  $10^{10}$ ($\circ$) and $t_w\!=\!10^{11}$ ($\triangleright$), $10^{13}$
  ($+$), $10^{15}$ ($\times$) in $d=3$. The solid lines have slope
  $\epsilon$ (small $t$ behavior) and slope $-\delta$ (large $t$
  behavior). For the value of the exponents $\epsilon$ and $\delta$,
  see text.}
\label{fig1}
\end{figure}
%************************************************************************
\noindent the trap model was so far considered only on a mean-field level
corresponding to an ``annealed situation'', where the site energies
are drawn anew after each jump.

In order to study aging effects in the glassy phase, we focus, as in
the trap model, on the (disorder averaged) probability
$\Pi(t_w\!+\!t,t_w)$ that the system does not change its state between
$t_w$ and $t_w+t$. For particles hopping on a lattice, this can be
interpreted as a dynamical structure factor \cite{Bouchaud/Dean:1995}.
Initially, the particle is located on any of the sites and then it
starts to explore the configuration space at some $T\!<\!T_{\rm g}$.
Physically, this means that we are considering an instantaneous quench
from $T=\infty$.

Our idea to explore the scaling properties of $\Pi(t_w\!+\!t,t_w)$ for
both $t$ and $t_w$ becoming large is based on the following ``partial
equilibrium'' concept: From the time after the quench up to $t_w$
the particle has followed a Brownian path in configuration space that
on average consists of $S(t_w)$ distinct and mutually connected sites.
On a typical path with $S\simeq S(t_w)$ sites then, it is reasonable
to think that the particles should effectively equilibrate, i.e.~the
probability to be on a particular site $j$ of the path may be
approximated by $\tau_j/\sum_{k=1}^S\tau_k$, where
$\tau_j\equiv\exp(-\beta E_j)$ ($1\!\le\!\tau_j\!<\infty$).
Conditioned on being at the site $j$, the system has a probability
$\exp(-t\sum_{n_j}w_{j,n_j})=\exp(-t\,\tau_j^{\alpha-1}\sum_{n_j}
\tau_{n_j}^\alpha)$ not to change state within time $t$, where the sum
over $n_j$ runs over all nearest neighbor sites of $j$.  Exactly two
of these neighboring sites are considered to belong to the Brownian
path in view of its one-dimensional topology. The remaining
$2(d\!-\!1)$ neighboring sites are assumed to have not been visited
before. Hence, we may deduce the scaling properties of
$\Pi(t_w\!+\!t,t_w)$ from
\begin{equation}
\tilde\Pi(t_w\!+\!t,t_w)\!\equiv\!\left\langle
\frac{\sum_{j=1}^{S(t_w)}\tau_j
    \exp\Bigl(-t\,\tau_j^{\alpha-1}\sum_{n_j}\hspace*{-0.1cm}
                                              \tau_{n_j}^\alpha\Bigr)}
     {\sum_{k=1}^{S(t_w)}\tau_k}
\right\rangle,
\label{pi1-eq}
\end{equation}
where $\langle\ldots\rangle$ denotes an average over $(2d-1)S(t_w)$
uncorrelated random numbers $\tau_j$ that are distributed according to
the power law $\phi(\tau)=\theta\tau^{-1-\theta}$ with
$\theta\!\equiv\!T/T_{\rm g}$.
 
Clearly, the partial equilibrium concept is an approximation that
needs to be tested. To this end we have determined
$\Pi(t_w+t,t_w)$ and $S(t_w)$ in $d\!=\!1,3$ for various
values of $\theta$ and $\alpha$ by Monte Carlo simulations. Then we
took $S(t_w)$ from these simulations to calculate
$\tilde\Pi(t_w\!+\!t,t_w)$ from eq.~(\ref{pi1-eq}). The disorder
average in this simple numerical evaluation was performed over a set
of realizations independent of the ones taken in the simulations.

Figure~\ref{fig1} shows the results for one representative parameter pair
$(\theta,\alpha)\!=\!(1/4,3/8)$. The data have been collected as functions of
$t$ for a broad range of fixed waiting times $t_w$ and are plotted already in
scaled form as functions of $\Lambda_1=t/t_w^{\mu_1}$ with exponents
$\mu_1\!=\!\mu_1(\theta,\alpha)$ being specified below. As can be seen from
the figure, the data for both $\Pi(t_w\!+\!t,t_w)$ and
$\tilde\Pi(t_w\!+\!t,t_w)$ collapse onto single curves
$F_1(\Lambda_1)$ and $\tilde F_1(\Lambda_1)$, respectively. Although the two
scaling functions are different, their asymptotic behavior for large
$\Lambda_1$ is the same, $F_1(\Lambda_1)\sim\tilde F_1(\Lambda_1)\sim
\Lambda_1^{-\delta}$ (see the solid lines in Fig.~\ref{fig1}). The values of
the exponent $\delta\!=\!\delta(\theta,\alpha)$ are given below.

It is important to inspect also more closely the `short' time regime,
where the complementary probability $1\!-\!\Pi(t_w\!+\!t,t_w)$ for the
system to change state between $t_w$ and $t_w\!+\!t$ is small. This
complementary probability can be as relevant as $\Pi(t_w\!+\!t,t_w)$
dependent on the physical quantity being measured. Scaling plots of
$1\!-\!\Pi(t_w\!+\!t,t_w)$ and $1\!-\!\tilde\Pi(t_w\!+t,t_w)$ as
functions of $\Lambda_1=t/t_w^{\mu_1}$ in $d\!=\!1,3$ are also shown
in Fig.~\ref{fig1}. Again there is a good data collapse and for
$\Lambda_1\to0$ we find $1\!-\!F_1(\Lambda_1)\sim 1\!-\!\tilde
F_1(\Lambda_1)\sim\Lambda_1^\epsilon$ with exponents
$\epsilon=\epsilon(\theta,\alpha)$ given below. An analogous overall
behavior as displayed in Fig.~\ref{fig1} was found also for other
pairs $(\theta,\alpha)$ (with $0\!<\theta\!<\!1$,
$0\!\le\!\alpha\!\le\!1$). We thus conclude that the partial
equilibrium concept not only yields the correct scaling behavior (same
$\mu_1$ exponents) but also the correct asymptotics of the scaling
functions \cite{dim-comm}. However, a study of the `participation ratios'
(see \cite{Compte/Bouchaud:1998} for their definition) shows that
the partial equilibrium concept is not exact, even for large
times \cite{Rinn/etal:2000}.

Now we turn to the analytical study of eq.~(\ref{pi1-eq}). Since
$\tilde\Pi(t_w\!+\!t,t_w)$ depends on $t_w$ only via $S(t_w)$ let us
first discuss the scaling of $S(t_w)$ with time $t_w$.  For
$\alpha\!=\!0$ this problem has been addressed some time ago (see
e.g.~\cite{Harder/etal:1987}) and for $t_w\!\to\!\infty$ one finds
\begin{equation}
S(t_w)\sim t_w^\gamma\,,
\hspace*{0.3cm}
\gamma=\left\{\begin{array}{cl}
\displaystyle\frac{d\,\theta}{d\!+\!(2\!-\!d)\theta}\,, & 1\le d<2 \\
\theta\,, & d>2 \end{array}\right.
\label{stw-eq}
\end{equation}
(In $d\!=\!2$ there are logarithmic corrections, $S(t_w)\sim
[t_w/\log t_w]^\theta$.)  For $0\!<\!\alpha\!\le\!1$ one
expects (\ref{stw-eq}) not to change, since $\alpha$ only affects the
nearest-neighbor hopping rates but not the transport properties on
large length scales. In fact, in our simulations we always found
eq.~(\ref{stw-eq}) to hold true. Let us note also that in $d\!=\!1$
one can give a simple finite-size scaling argument to show that
(\ref{stw-eq}) remains valid for $0\!<\!\alpha\!\le\!1$.

Next we derive the scaling properties of $\tilde\Pi$ as a function of
$t$ and $S=S(t_w)$, and then use eq.~(\ref{stw-eq}) to obtain the
corresponding scaling properties of $\tilde\Pi$ as a function of $t$
and $t_w$.  When replacing the denominator in eq.~(\ref{pi1-eq}) by
$\int_0^\infty d\lambda\, \exp(-\lambda\sum_{k=1}^S\tau_k)$, the
average over the $\tau_j$ can to a large extent be factorized, and one
obtains the following asymptotic formula valid in the limit of large
$S$
\begin{eqnarray}
\tilde\Pi&\cong&\frac{\theta^3\tilde S}{\kappa}
\int_0^\infty\hspace*{-0.2cm}d\lambda\, e^{-\lambda^\theta \tilde S}
\hspace*{-0.1cm}
\int_1^\infty\hspace*{-0.1cm}
        \frac{d\tau_1}{\tau_1^{1+\theta}}\,e^{-\lambda\tau_1}
\hspace*{-0.1cm}
\int_1^\infty\hspace*{-0.1cm}
       \frac{d\tau_2}{\tau_2^{1+\theta}}\,e^{-\lambda\tau_2}
\nonumber\\
&&\hspace*{0cm}\times\int_1^\infty\hspace*{-0.1cm}
\frac{d\tau}{\tau^\theta}\,e^{-\lambda\tau}
\Bigl[f\bigl(\frac{t}{\tau^{1-\alpha}}\bigr)\Bigr]^{2(d\!-\!1)}
\exp\Bigl(-t\frac{\tau_1^\alpha\!+\!\tau_2^\alpha}{\tau^{1-\alpha}}\Bigr)\,,
\nonumber\\
&&f(x)\equiv\theta\int_1^\infty \frac{d\tau}{\tau^{1\!+\!\theta}}\,
\exp(-x\tau^\alpha)\,.
\label{pi2-eq}
\end{eqnarray}
Here $\kappa\equiv\Gamma(1\!-\!\theta)$, where $\Gamma(.)$ is the
Gamma function, and $\tilde S\equiv \kappa S$. Note that for $t\!=\!0$
and $\tilde S\!\to\!\infty$, eq.~(\ref{pi2-eq}) yields the correct
normalization $\tilde\Pi\!\to\!1$.

After two transformations $\lambda\!\to\!\tilde S\lambda^\theta$ and
$\tau\!\to\!\tilde S^{1/\theta}\tau$ we can identify $\Lambda_1\!=\!t/\tilde
S^{(1\!-\!\alpha)/\theta}$ as a scaling variable corresponding to a first
characteristic time $t_1\!\sim\!\tilde S(t_w)^{(1\!-\!\alpha)/\theta}\sim
t_w^{\gamma(1\!-\!\alpha)/\theta}\equiv t_w^{\mu_1}$.  We thus obtain
\begin{equation}
\mu_1=\frac{\gamma(1-\alpha)}{\theta}
\label{mu1-eq}
\end{equation}
with $\gamma$ from eq.~(\ref{stw-eq}). This exponent $\mu_1$ has been
used to collapse the data in Fig.~\ref{fig1}, i.e. we took
$\mu_1=(1\!-\!\alpha)/(1\!+\!\theta)$ in $d\!=\!1$ and
$\mu_1=(1\!-\!\alpha)$ in $d\!=\!3$. Based on the equilibrium concept
it is is easy to show that $t_1$ has a simple physical interpretation:
It scales as the typical maximum trapping time $t_{\max}(t_w)$
encountered after $t_w$. In $d\!=\!1$ one thus finds a subaging
behavior even for $\alpha\!=\!0$ since the deepest trap is visited a
large number of times $N(t_w)\!\sim\!t_w^{\theta/(1\!+\!\theta)}$, so
that $t_1\!\sim\!t_{\max}(t_w)\!\sim\!t_w/N(t_w)\!\ll\!t_w$.
Similarly, for $\alpha\!\ne\!0$, the deepest trap is revisited a large
number of times in all dimensions $d$ due to a strong backward jump
correlation when the particle leaves a site with low energy.

The scaling function in the first time domain $t\!=\!\Lambda_1 t_w^{\mu_1}$
reads
\begin{eqnarray}
\tilde F_1(\Lambda_1)&=&\frac{\theta^3}{\kappa}
\int_0^\infty\hspace*{-0.2cm}d\lambda\, e^{-\lambda^\theta}\hspace*{-0.1cm}
\int_1^\infty\hspace*{-0.1cm}
        \frac{d\tau_1}{\tau_1^{1+\theta}}
\hspace*{-0.1cm}
\int_1^\infty\hspace*{-0.1cm}
       \frac{d\tau_2}{\tau_2^{1+\theta}}\\
&&\hspace*{-0.8cm}\times\int_0^\infty\frac{d\tau}{\tau^\theta}e^{-\lambda\tau}
\Bigl[f\bigl(\frac{\Lambda_1}{\tau^{1-\alpha}}\bigr)\Bigr]^{2(d\!-\!1)}
\exp\Bigl(-\Lambda_1
   \frac{\tau_1^\alpha\!+\!\tau_2^\alpha}{\tau^{1-\alpha}}\Bigr)\,,\nonumber
\end{eqnarray}
%************************************************************************
\begin{figure}
\hspace*{-0.5cm}\epsfig{file=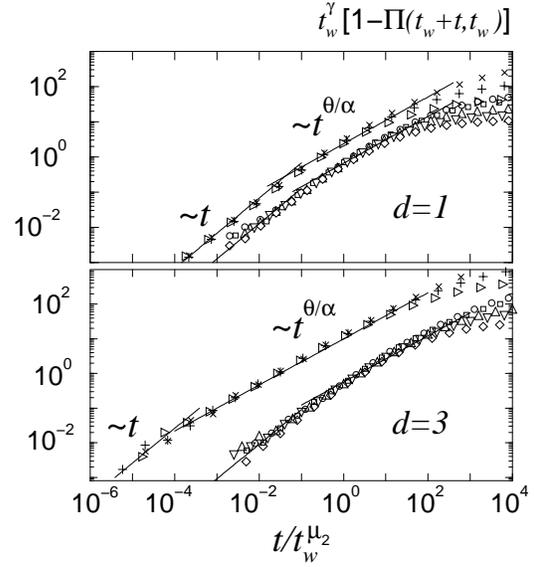,width=8cm}
\caption{Double-logarithmic plot of
  $1\!-\!\Pi(t_w\!+\!t,t_w)$ and $1\!-\!\tilde\Pi(t_w\!+\!t,t_w)$ as
  functions of $t/t_w^{\mu_2}$ in $d\!=\!1$ and $d\!=\!3$ for the same
  parameters as in Fig.~\ref{fig1} (the same symbols have been used
  for the various waiting times $t_w$). The solid lines have slope one
  and $\theta/\alpha$, and indicate the limiting behavior of $F_2(.)$
  according to eq.~(\ref{f2lim-eq}).}
\label{fig2}
\end{figure}
%************************************************************************
\noindent and has the limiting behavior
\begin{equation}
\tilde F_1(\Lambda_1)\sim\left\{\begin{array}{lll}
1-c_{\scriptscriptstyle>} \Lambda_1^{(1\!-\!\theta)/(1\!-\!\alpha)}\,, 
         & \theta>\alpha & \\[-0.15cm]
& & \hspace*{-0.5cm}\Lambda_1\to0 \\[-0.15cm]
1-c_{\scriptscriptstyle<} \Lambda_1^{\theta/\alpha}\,, 
         & \theta<\alpha & \\[0.25cm]
c_{\scriptscriptstyle\infty} \Lambda_1^{-\theta/(1\!-\!\alpha)}
         & \hspace*{0.5cm} \Lambda_1\to\infty & \end{array}\right.
\label{f1lim-eq}
\end{equation}
The constants $c_{\scriptscriptstyle>}$, $c_{\scriptscriptstyle<}$, and
$c_{\scriptscriptstyle\infty}$ can be expressed in terms of $\alpha$, $\theta$
and $d$ but are not of interest here.  Equation~(\ref{f1lim-eq}) yields the
exponents $\delta\!=\!\theta/(1\!-\!\alpha)$ and
$\epsilon\!=\!(1\!-\!\theta)/(1\!-\!\alpha)$ taken in Fig.~1 to characterize
the decay of $\Pi$ and rise of $1\!-\!\Pi$, respectively. For $\alpha=0$, one
recovers the results of the annealed model \cite{Bouchaud/Dean:1995}, namely
$\delta\!=\!\theta$ and $\epsilon\!=\!1\!-\!\theta$.

For $\theta\!>\!\alpha$ the function $\tilde F_1(\Lambda_1)$ describes
the scaling properties of $\tilde\Pi$ completely. However, based on
eq.~(\ref{pi2-eq}) one finds that for $\theta\!<\!\alpha\!<\!1/2$,
there exists a second scaling variable $\Lambda_2\!=\!t/\tilde
S^{(1\!-\!2\alpha)/\theta}\sim
t/t_w^{\gamma(1\!-\!2\alpha)/\theta}\equiv t/t_w^{\mu_2}$ yielding
\begin{equation}
\mu_2=\frac{\gamma(1-2\alpha)}{\theta}\,.
\label{mu2-eq}
\end{equation}
Therefore, a {\it second} characteristic time
scale $t_2\!\sim\!S^{(1\!-\!2\alpha)/\theta}$, diverging 
when $t_w \to \infty$, governs the behavior when $\tilde\Pi$ is close to
one. (Note that for fixed $\Lambda_2$,
$\Lambda_1\!=\!\tilde S^{-\alpha/\theta}\Lambda_2\!\to\!0$ for
$\tilde S\!\to\!\infty$). In the new time domain one finds the 
following {\it generalized scaling} form,
\begin{equation}
\tilde S(1-\tilde\Pi)=\tilde F_2(\Lambda_2)\,,\hspace*{0.5cm} 
\theta\!<\!\alpha\!<\!1/2\,,
\label{genscal-eq}
\end{equation}
where $F_2(.)$ is given by
\begin{eqnarray}
\tilde F_2(\Lambda_2)&=&
\frac{\psi_\infty\theta^2{\Lambda_2}}{(1\!-\!\alpha)\kappa}
\int_0^\infty\hspace*{-0.12cm}
   \frac{d\lambda\, e^{-\lambda^\theta}}{\lambda^{\alpha-\theta}}
\hspace*{-0.15cm}
\int_0^\infty\hspace*{-0.12cm}
\frac{d\tau\, e^{-\lambda\tau}}{\tau^{1-\alpha+x}}\,
g(\Lambda_2\lambda^{1\!-\!\alpha}\tau^\alpha),\hspace*{-1cm}\nonumber\\
&&\hspace*{-1.5cm}g(x)\equiv x^{\frac{\alpha-\theta}{1-\alpha}}
\hspace*{-0.15cm}
\int_0^\infty \hspace*{-0.25cm}\frac{du}{u^{1+\frac{1-\theta}{1-\alpha}}}
(1\!-\!e^{-u})\,\exp[-(x/u)^{\frac{1}{(1-\alpha)}}].
\label{f2-eq}
\end{eqnarray}
Here $\psi_\infty\!\equiv\!
\lim_{\bar\tau\to\infty}\psi(\bar\tau)/(\theta\bar\tau^{-1\!-\!\theta})$, and
$\psi(\bar\tau)$ denotes the probability distribution for the variable
$\bar\tau\!\equiv\!(\tau_1^\alpha\!+\!\tau_2^\alpha)^{1/\alpha}$, i.e.
$\psi(\bar\tau)\!\equiv\!\langle
\delta(\bar\tau\!-\![\tau_1^\alpha\!+\!\tau_2^\alpha]^{1/\alpha})\rangle$.
From (\ref{f2-eq}) follows
\begin{equation}
\tilde F_2(\Lambda_2)\sim\left\{\begin{array}{ll}
c_0\, \Lambda_2\,, & \Lambda_2\to0  \\[0.15cm]
c_{\scriptscriptstyle<}\,\Lambda_2^{\theta/\alpha}\,,
         & \Lambda_2\to\infty  \end{array}\right.
\label{f2lim-eq}
\end{equation}
where $c_0$ is a constant dependent on $\theta$ and $\alpha$. We note
that $1\!-\!\tilde\Pi$ matches continuously as one leaves the short
time scaling regime ($t\!\sim\!t_w^{\mu_2}$) described by $F_2$ to
enter the regime described by the scaling function $F_1$ (where
$t\!\sim\!t_w^{\mu_1}$). Figure 2 shows the short time behavior of
$\tilde\Pi$ and $\Pi$, rescaled as in (\ref{genscal-eq}) for the same
parameters as in Fig.~\ref{fig1}. As can be seen from the figure, the
data approach the two power laws predicted by eq.~(\ref{f2lim-eq}) for
large $t_w$.

For clarity, we illustrate the overall behavior of
$\Pi(t_w\!+\!t,t_w)$ as a function of $t$ in Fig.~3. For
$(\theta,\alpha)$ values in the two-time-scaling region
$0\!<\!\theta\!<\!\alpha\!<\!1/2$ (shaded area of the
$\alpha$-$\theta$-diagram shown in the inset), there exist three
different $t$ regimes: ({\it I}\,) $\Pi(t_w\!+\!t,t_w)\!\sim\!
1\!-\!{\rm const.}\,t_w^{-\gamma}[t/t_w^{\mu_2}]$ for $t\!\ll\!
t_w^{\mu_2}$, ({\it II}\,) $\Pi(t_w\!+\!t,t_w)\!\sim\! 1\!-\!{\rm
  const.}'\, t_w^{-\gamma}[t/t_w^{\mu_2}]^{\theta/\alpha}=1\!-\!{\rm
  const.}'\, [t/t_w^{\mu_1}]^{\theta/\alpha}$ for
$t_w^{\mu_2}\!\ll\!t\!\ll\!t_w^{\mu_1}$, and ({\it III}\,)
$\Pi(t_w\!+\!t,t_w)\!\sim\! [t/t_w^{\mu_1}]^{-\delta}$ for
$t_w^{\mu_1}\!\ll\!t$. When $(\theta,\alpha)$ lies in the unshaded
area of the $\alpha$-$\theta$-diagram, the first regime
$t\!\ll\!t_w^{\mu_2}$ does not exist (or, more precisely, it then
becomes irrelevant in the limit of large $t_w$). Note that for
$t_w\!\to\!\infty$ and $\Lambda_2\!=\!t/t_w^{\mu_2}$ fixed, the
long-time regime in Fig.~3 ``moves toward infinity'' and the behavior
is fully described by the second scaling function $F_2(\Lambda_2)$,
while for $t_w\!\to\!\infty$ and $\Lambda_1\!=\!t/t_w^{\mu_1}$ fixed
the short-time regime in Fig.~3 ``moves toward zero'', and the
behavior is fully described by the first scaling function
$F_1(\Lambda_1)$.

In summary we have shown {\it (i)} that generalized trap models can
exhibit subaging behavior, induced by multiple visits to the same
trap, and {\it (ii)} the possible existence
%************************************************************************
\begin{figure}
\hspace*{1cm}\epsfig{file=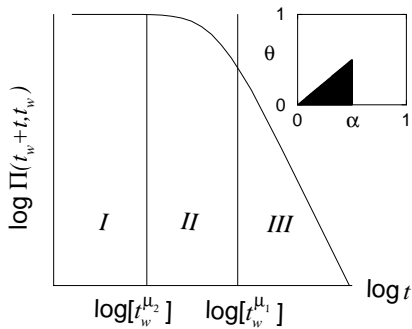,width=5.6cm}
\vspace*{-0.5cm}
\caption{Sketch of the behavior of $\Pi(t_w\!+\!t,t_w)$
  as a function of time $t$ in the three regimes ({\it I}-{\it
    III}\,).  The shaded area in the $\alpha$-$\theta$-diagram marks
  the two-time scaling region.}
\label{fig3}
\end{figure}
%************************************************************************
\noindent of several distinct
scaling regimes in the two-time plane. Such a possibility is of
crucial importance for the interpretation of experiments, since the
waiting time can typically be varied between one minute and a few days
only (with some notable exceptions \cite{Struik:1978}). The occurrence
of several time regimes then may get masked by an apparent rescaling
of the relaxation curves by a single effective value of $\mu$. From a
theoretical perspective, it would be interesting to study the
existence of a generalized Fluctuation-Dissipation Theorem in the
aging regime, as it is predicted by mean-field spin-glass models
\cite{Cugliandolo/etal:1997}. This problem is intimately related to
the validity of the partial equilibrium concept introduced above.

We should like to thank W.~Dieterich and E.~Pitard for stimulating
discussions. P.M. gratefully acknowledges financial support from the
Deutsche Forschunggsgemeinschaft (Ma~1636/2-1).

\end{multicols}


\begin{references}
\bibitem{dim-comm} One might object that the partial equilibration
  concept ceases to be valid above some critical dimension $d_c$.
  However, for $(\theta,\alpha)\!=\!(1/3,1/4)$ we tested the concept
  in $d\!=\!10$ and $d\!=\!100$ also, and found it to hold true.
\end{references}
\end{document}